# Short remarks on possible production of defects in low temperature semiconductor detectors for dark matter physics experiments

Ionel Lazanu[1] and Sorina Lazanu[2]

[1] University of Bucharest, Faculty of Physics, POBox MG-11, Bucharest-Măgurele, Romania
[2] National Institute of Materials Physics, POBox MG-7, Bucharest-Măgurele, Romania

Emails: ionel.lazanu@g.unibuc.ro and lazanu@infim.ro

**Abstract** The nature of dark matter is still an open problem, but there is evidence that a large part of the dark matter in the universe is non-baryonic, non-luminous and non-relativistic and hypothetical Weakly Interacting Massive Particles (WIMPs) are candidates that satisfy all of the above criteria. In order to minimize the ambiguities in the identification of WIMPs' interactions in their search, in more experiments, two distinct quantities are simultaneously measured: the ionization and phonon or light from scintillation signals. Silicon and germanium crystals are used in some experiments. In this paper we discuss the production of defects in semiconductors due to WIMP interactions and estimate their contribution in the energy balance. This phenomenon is present at all temperatures, is important in the range of keV energies, but is not taken into consideration in the usual analysis of experimental signals and could introduce errors in identification for WIMPs.

## 1. Introduction

The search for dark matter has become a very active research area in the last decades. There is evidence that a large part of the dark matter in the universe is non-baryonic, non-luminous and non-relativistic. Hypothetical Weakly Interacting Massive Particles (WIMPs) are proposed as particle candidates that satisfy all of the above criteria.

Thus, the detection of low-energy nuclear recoils originating from WIMPs interactions (or from neutrinos, the other candidates) is the subject of a great experimental effort.

The difficulties involved in the development of detectors adequate for this task are multiple because the signal rate expected is generally very low (few counts per kg target per day or much less), the signal is in the range of keV or lower, requiring a low energy threshold, and this signal must be extracted from a dominant background.

In more recent experiments, the ionization and phonon (or light from scintillation signals) are measured simultaneously, trying to discriminate both between electrons - nucleon/nuclei recoils and also between different sources of the process: ordinary matter or constituents of the dark matter. In collaborations as e.g. CDMS, EDELWEISS, or GENIUS, semiconductor crystals of silicon and germanium are used.

In this paper we discuss and estimate the possible errors in the identification of WIMPs due to the production of defects in semiconductors, process present at all temperatures, important in the range of keV energies, but that is not taken into consideration in the usual analysis of experimental signals [1].

## 2. Physical processes and their modelling
*2.1 WIMP direct detection*
There are many experimental observations suggesting that the largest part of the matter of the Universe should consist in weakly interacting, non-baryonic, massive and stable particle. WIMPs should be gravitationally trapped inside the galaxy and their mean velocity inside our galaxy is expected to be in the order of hundreds of km/s. It should be possible to detect WIMPs by their elastic scattering with the nuclei of ordinary matter ("direct detection") where the mass of WIMPs is in the range GeV up to 1 TeV. At those velocities, WIMPs should produce typical nuclear recoil energies in the range of tens of keV.

Supposing for WIMPs an average velocity of the order of 230 km/s [2] and elastic interactions in detector, the energies transferred to recoil nuclei have been calculated. The dependence of the recoil energy of the primary knock-on nucleus in silicon, as a function of WIMP mass and scattering angle is represented in figure 1.

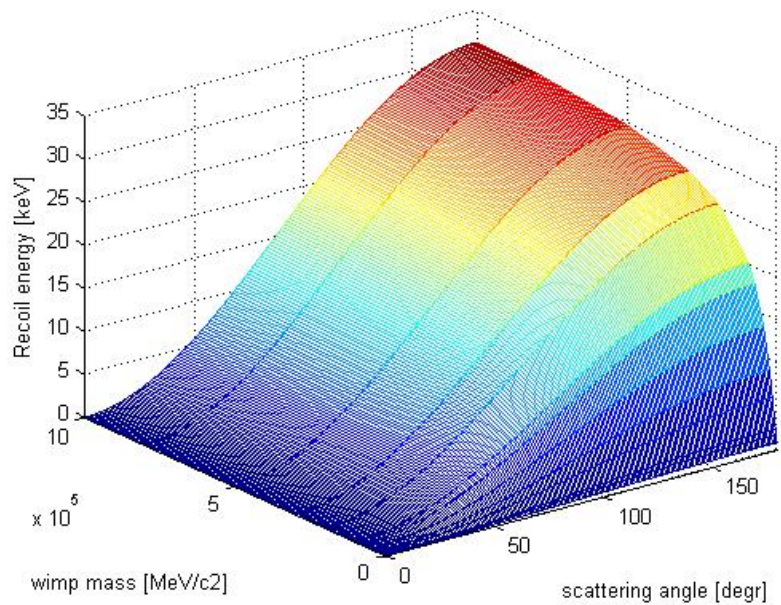

Figure 1. Energy distribution of nuclear recoils in silicon versus WIMP mass and scattering angle for average WIMP velocity of 230 km/s.

Most probably, the WIMP interacts in a singular interaction, producing a nuclear recoil that initiates a cascade of new interactions.

*2.2 Cascade of interactions and defect production by nuclear recoils*
The energy lost by the nuclear recoil in the semiconductor target is divided between ionization (communicated to the electron system), phonons (oscillations of the nuclei in the sites of the crystal) and creation of lattice defects. Both the energy deposited into defect creation and that transmitted as ionization could partially be transferred to the phonon system by defect annihilation and by recombination of electron-hole pairs. In fact, in calorimetric conditions the energy balance is: $E_{recoil} = E_{ionization} + E_{phonon} + E_{stable\ defects}$. The evaluation of the quantity of energy stored in "stable" defects, after the transitory regime, is a difficult task.

The mechanism of primary defect production works equally well at very low temperatures. It could be described by the following succession of steps: the incident particle hits a nucleus of the target, the primary knock-on, and, if its energy is higher than the displacement energy $E_d$, it displaces it from its site in the lattice, leaving behind a vacancy. If the transmitted energy is lower than the threshold energy for displacements, this energy is transferred to the phonon system. On the other side, if the energy of the PKA is high enough, it creates a displacement cascade, composed of equal numbers of vacancies and interstitials, which could also be in the form of close Frenkel pairs. Another type of primary defect, both in silicon and in germanium, is the four-folded coordination defect (FFCD) [3 ÷ 6].

The vacancy and the interstitial destroy the fourfold coordination of the lattice and relatively high defect formation energy for these defects is the consequence. For both silicon and germanium it is in the order of 3 ÷ 6 eV [4, 7, 8]. The Frenkel pair is also a defect that conserves the number of particles, and its formation energy is less than the sum of an isolated vacancy and interstitial. In contrast to all these point defects, in the FFCD only two bonds are broken, the formation energy is lower in respect to previously mentioned defects, and the bond length and angles do not significantly deviate from their bulk values.

In order for a permanent defect to be produced from an initially perfect crystal lattice, the kinetic energy that it receives must obviously be larger than the formation energy of a Frenkel pair. However, while the Frenkel pair formation energies in crystals are typically around 5–10 eV, the average threshold displacement energies are much higher, in the order of 20 eV or higher. The reason for this apparent discrepancy is that the defect formation is a complex multi-body collision process (a small collision cascade) where the atom that receives recoil energy can also bounce back, or kick another atom back to its lattice site. Hence, even the minimum threshold displacement energy is usually clearly higher than the Frenkel pair formation energy. In fact, the displacement energy is the sum between the energy of formation of vacancy ($\Delta H_f^V$) and interstitial ($\Delta H_f^I$), and a quantity which goes to the lattice ($E_L$), consisting mostly in a bond-bending component [9]:

$$E_d = \Delta H_f^V + \Delta H_f^I + E_L \tag{1}$$

The concrete values for $E_d$ represent a controversial problem [10], because their value depend on crystal symmetry and orientation, direction of the recoil in the lattice (which is function of the direction of the irradiation beam in respect to the axes of symmetry of the crystal, energy and mass of incident particle) and temperature.

At temperatures around 1K and under this value, the main difference in respect to higher temperatures is related to the mobility of defects: Si and Ge must be discussed separately.

For the case of low temperature irradiation of silicon, vacancies are frozen in, and close Frenkel pairs are stable at sub-Kelvin temperatures. Interstitials have athermal migration by the Bourgoin - Corbett mechanism [11], and have never been evidenced experimentally. Single vacancies and impurity interstitials (as, e.g. $Al_i$, $B_i$) [12, 13] were put in evidence. The presence of radiation-produced close Frenkel pairs, stable at the irradiation temperature, has been detected by X-ray measurements on Si irradiated with fast electrons at liquid helium temperature [14]. These defects anneal out after some authors in the temperature range 10 – 70K [15] or, after others (Refs. [16, 17]), around 140 – 160 K.

In contrast with silicon, interstitials in germanium could be observed; information about the properties of isolated *I* and *V* in Ge has first been obtained by perturbed angular correlation spectroscopy (PAC) trapping experiments [18], as well as by DLTS [19], following low temperature electron irradiation. It seems that the Frenkel pair in *n*-type germanium could be frozen into the lattice at temperatures below 65K, and thus observed [20]. Germanium self-interstitials and probably vacancies disappear at around 200 K. In *p*-type germanium, the main defects observed are the single vacancy and the gallium interstitial [19]. Both disappear after reaching room temperature without converting to any other defect, such as divacancy *V-V* or $Ga_i-V$, as is the case in silicon.

The marked difference between Si and Ge is the fact that in Si, primary defects interact with other defects and impurities, as well as between themselves at all temperatures, producing "complex" defects, while in (p-type) Ge the annealing could be complete.

The average number of displaced atoms in the cascade produced by a PKA of energy $E$ has been estimated using the modified Kinchin - Pease displacement damage function [21 ÷23]. The basic assumptions used in its derivation are that the collisions are binary, elastic and made between similar atoms (implying that if the energy is less than $E_d$, than there is not displaced atom, while atoms receiving energy between $E_d$ and $2E_d$ are displaced from their lattice position, but do not contribute to any additional displacements due to subsequent collisions); a correction factor $\xi$ is applied for PKA energies greater than $2E_d$, to account for the electronic stopping power of the material. Therefore, the displacement damage function is:

$$<N_d> = \begin{cases} 0 & E < E_d \\ 1 & E_d \leq E < 2E_d \\ \dfrac{\xi E}{2E_d} & 2E_d \leq E < E_c \\ \dfrac{\xi E_c}{2E_d} & E \geq E_c \end{cases} \quad (2)$$

where $E_c$ is a cutoff energy, of the order of 28 keV for Si targets and 73 keV for Ge ones [24].

Molecular dynamics simulations of collision cascades initiated by PKA ranging from 100 eV to 1 keV [25] put in evidence a solid to liquid like transformation at the height of the cascade. Upon cooling, the liquid-like regions collapse, resulting in the formation of numerous isolated defects and clusters of defects, collision events following the modified Kinchin Pease model. These simulations found almost no temperature dependence of the collision events at short times, of the order of tens of ps.

The numbers of vacancy-interstitial pairs created by a single WIMP elastic interaction in silicon and germanium lattices have been calculated, supposing threshold displacement energies of 21 [26] and 25 eV [27] respectively. Figures 2 and 3 present the results for Si and Ge respectively. It could be seen that at high WIMPs masses, the number of vacancy-interstitial pairs is higher in Ge; this is mainly due to the higher energy transferred to the lattice recoil in the elastic interaction. Equal energies transferred to silicon and germanium nuclei correspond to a WIMP mass of around 50 GeV/c$^2$. The process of creation of vacancy – interstitial pairs is the most energy consuming from the possible processes of production of primary defects mentioned above, and this is the reason it was considered here.

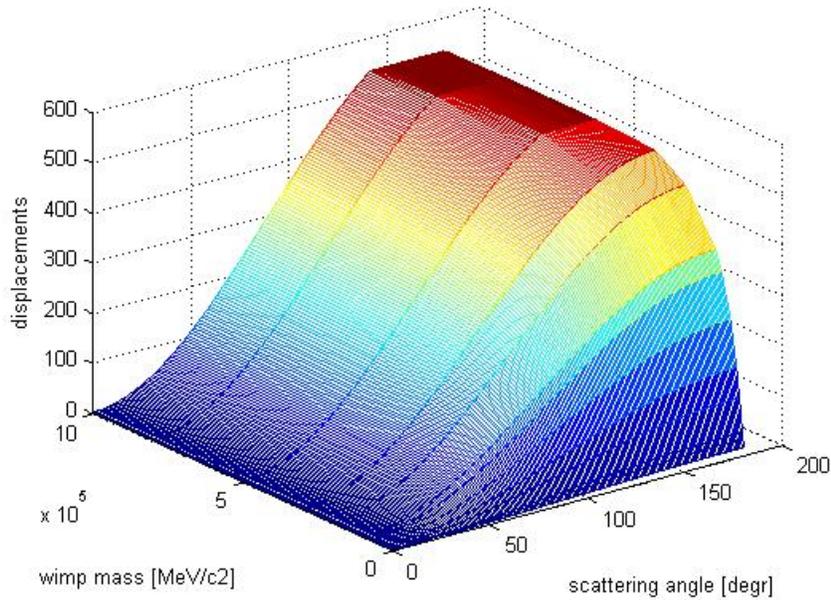

Figure 2: Displacement damage function in Si Vs. WIMP mass and scattering angle

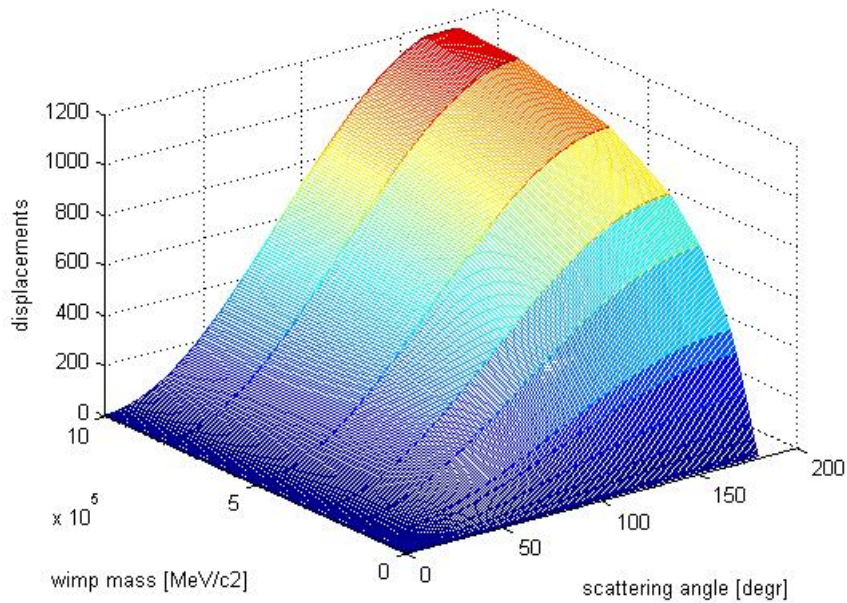

Figure 3 Displacement damage function in Ge Vs. WIMP mass and scattering angle

## 3. Results and comments

An estimation of the recombination rate of vacancies and interstitials at very low temperatures is obtained in this paper from the measured values of the introduction rates of defects reported in the literature. The introduction rate of defects in Si has been determined based on different types of measurements, e.g. from the maximum in ac hopping conductance [28], change in the Schottky barrier capacitance [13], investigations of the diffuse scattering intensity and of the change of the lattice parameter [29] following MeV electron irradiation. The compilation of defect introduction rates at low temperatures includes references [13, 15, 25, 28÷30]. The introduction rates depend on the material and on its doping, but have weak temperature dependence. Introduction rates of the order of 1 cm$^{-1}$ are typical. A survival rate of vacancy-interstitial pairs in silicon of ~ 80% has been estimated and is used in the present calculations. Supposing that the modified Kinchin – Pease model reproduces correctly the production of defects, considering the same survival rate of defects in both silicon and germanium, and taking into consideration the energy balance (1), we estimated a maximal limit of around 12% of recoil energy deposited into defects in silicon and 14% for Ge, that introduces an error in the mass identification of WIMPs considering the kinematics of the interaction up to 25 % in silicon and up to 40 % in germanium respectively..

## 4. Summary and conclusions

In this paper, the contribution of production of defects in semiconductor detectors due to WIMPs interactions was estimated. This represents a source of errors in WIMPs mass identification.

A difficulty in the evaluation of the energy stored in defects is represented by the fact that more types of primary defects are possible: in this respect, the most unfavourable case was considered, when only vacancy – interstitial pairs were formed. Another difficulty is related to the values used both for the displacement energy, and for the energy of formation of vacancy and interstitial, most of them with a wide scatter. The correlation between the degradation and macroscopic changes in detector parameters or the theoretical modelling of degradation in the semiconductor calorimeter represent possible ways to correct the errors in identification.

**Acknowledgements**
This work was supported by the Romanian National University Research Council, under contract IDEI 901/2008.